\documentclass[prl,twocolumn,showpacs,amsmath,amssymb,superscriptaddress]{revtex4}
\usepackage[latin1]{inputenc}
\usepackage{graphicx}
\usepackage{dcolumn}
\usepackage{bm}
\begin{document}
\title{Current percolation and anisotropy in polycrystalline MgB$_2$}
\author{M.~Eisterer}
\email{eisterer@ati.ac.at} \affiliation{Atominstitut der
Österreichischen Universitäten, A-1020 Vienna, Austria}
\author{M.~Zehetmayer}
\affiliation{Atominstitut der Österreichischen Universitäten,
A-1020 Vienna, Austria}
\author{H.~W.~Weber}
\affiliation{Atominstitut der Österreichischen Universitäten,
A-1020 Vienna, Austria}
\date{\today}
\begin{abstract}
The influence of anisotropy on the transport current in  MgB$_2$
polycrystalline bulk samples and wires is discussed. A model for
the critical current density is proposed, which is based on
anisotropic London theory, grain boundary pinning and percolation
theory. The calculated currents agree convincingly with
experimental data and the fit parameters, especially the
anisotropy, obtained from percolation theory agree with experiment
or theoretical predictions.
\end{abstract}

\maketitle

Soon after the discovery of superconductivity in MgB$_2$
\cite{Nag01}, it was found that grain boundaries do not limit the
current flow in polycrystalline samples \cite{Lar01,Kam01}. Thus,
sintered bulk samples and wires could be prepared with critical
current densities appropriate for technical applications at low
magnetic fields \cite{Buz01}. With increasing field the critical
current rapidly decreases, becoming zero at a field far below the
upper critical field. Anisotropy, observed on thin films
\cite{Pat01} for the first time, was identified as a possible
reason for this strong field dependence \cite{Chr02}. The aim of
this letter is to evaluate the influence of anisotropy on the
field dependence of $J_c$ in terms of percolation theory
\cite{Sta92,Kir73}. Percolation theory had been employed
frequently to analyze currents in granular superconductors
\cite{Dav76,Deu80,Ent84,Spe96,Pre96,Osa00,Has00,Nak02,Zei02},
especially in high temperature superconductors, where grain
boundaries limit the current flow, and also to interpret the
irreversibility line in MgB$_2$ coatings \cite{Chr02}. We present
a percolation model for the limitation of the transport current by
anisotropy. While some details refer to MgB$_2$, the model is
applicable to any granular anisotropic superconductor or, more
generally, to other systems with spatially inhomogenous transport
properties.

The superconductor is assumed to consist of randomly oriented
grains with exactly  identical properties. Thermal activation of
the flux line lattice is neglected, since its influence in
granular MgB$_2$ should not be much larger than in the low-T$_c$
materials NbTi or Nb$_3$Sn \cite{Pat01} (this simplification might
be the reason for some systematic deviations at high temperatures,
cf. below). At fixed temperature the upper critical field only
depends on the angle between the boron layers in the grain and the
applied magnetic field according to the anisotropic Ginzburg
Landau relation \cite{Til65}
\begin{equation}
B_{c2}(\theta)=\frac{B_{c2}(\pi/2)}{\sqrt{\gamma^{2}\cos^{2}(\theta)+\sin^{2}(\theta)}}
\label{equ1}
\end{equation}
$\gamma$ denotes the anisotropy factor of the upper critical
field, i.e. $B_{c2}(\pi/2)/B_{c2}(0)=:B_{c2}^{\|}/B_{c2}^{\bot}$,
and $\theta$ is the angle between the applied field and the
c-axis. This relation was confirmed experimentally in MgB$_2$ by
torque measurements \cite{Ang02,Zeh02}. We start by considering
the resistive transition in decreasing temperature at fixed field
$B_{0}$. At a certain temperature, $B_{c2}^{\|}$ becomes equal to
the applied field $B_{0}$ and the resistivity starts to decrease.
This point is per definition $B_{c2}$ of the whole sample. When
the temperature decreases further, an increasing number of grains
becomes superconducting, depending on their orientation. During
this transition the system can be considered as a mixture of
normal and superconducting grains. Systems consisting of two
materials with different resistances (conducting/insulating
\cite{Las71,Abe75,Cle80,Lee86} or normal/superconducting
\cite{Her84,Xia88}) have been investigated experimentally and are
well understood in the framework of percolation theory. In
mixtures of normal and superconducting powders the resistivity
remains finite as long as the probability $p$, that a grain is
superconducting (i.e. the fraction of superconducting grains), is
smaller than the critical probability or percolation threshold
$p_{c}$. The percolation threshold strongly depends on the number
of connections to neighboring grains and, therefore, on the
preparation conditions. When the fraction of superconducting
particles becomes equal to the critical probability, the first
continuous superconducting current path occurs and the resistivity
disappears for sufficiently small currents. Since the upper
critical field increases monotonically with $\theta$, an angle
$\theta_{c}$ corresponds to the critical probability within the
transition region. If the boron planes are randomly oriented, the
directions of the c-axes are equally distributed in 3D space and
their angles to any fixed direction are distributed as
$\sin(\theta)$. Integration leads to $\theta_{c}=\arccos(p_{c})$.
With $B_{c2}(\pi/2)=B_{0}$ and $B_{c2}(\theta_{c})=B_{0}$ for the
onset and the offset of the transition, and assuming
$\frac{\partial B_{c2}}{\partial T}$ and $\gamma$ to be constant
within the transition region, the transition width is obtained
from (\ref{equ1}):
\begin{equation}
\triangle
T_{a}=\frac{\sqrt{(\gamma^{2}-1)p_{c}^{2}+1}-1}{(-\frac{\partial
B_{c2}}{\partial
T})}B_{0}=:\frac{f_{tw}(\gamma,p_{c})}{(-\frac{\partial
B_{c2}}{\partial T})}B_{0} \label{equ2}
\end{equation}
$\triangle T_{a}$ represents the broadening of the transition only
due to the anisotropy and is, therefore, additive to other reasons
for a finite transition width, e.g. material inhomogeneities,
thermal fluctuations, thermal gradients in the sample, surface
effects etc. The numerator of (\ref{equ2}), $f_{tw}$, can be
expressed explicitly and interpreted as a normalized transition
width. It is a monotonically increasing function of $\gamma$ and
$p_{c}$.

\begin{figure}
\centering \includegraphics[width = \columnwidth]{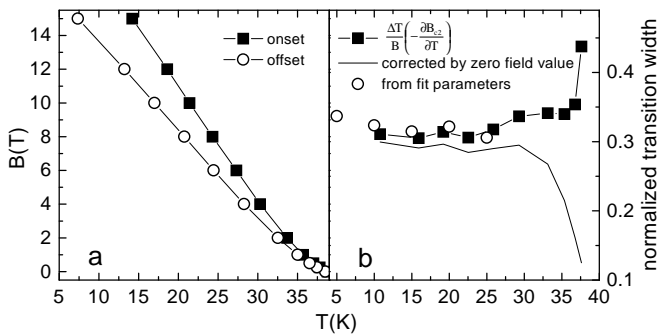}
\caption{Onset and offset (a) of the resistive transition in bulk
MgB$_2$ and normalized transition width (b).} \label{fig1}
\end{figure}

The resistive transition of a sintered bulk sample \cite{Kam01}
was measured at fixed magnetic field at a current density of 2
kA/m$^{2}$.  The onset of the transition was defined at 95 \% of
the normal state resistance, the offset at 5 \%. The results are
plotted in Fig.~\ref{fig1}a. From the difference of onset and
offset temperature, $f_{tw}$ is calculated and plotted in
Fig.~\ref{fig1}b (black squares) as a function of the mean value
of these two temperatures. Since the contribution of anisotropy
tends to zero at zero field, it cannot be dominant at low fields,
which explains the upturn of the normalized transition width at
high temperatures. As a simple correction, we subtract the zero
field transition width (0.25 K) at all fields, which leads to the
solid line (without symbols) in Fig.~\ref{fig1}b. The upturn of
the original data is removed and $f_{tw}$ becomes nearly
temperature independent below ~30 K, but decreases rapidly at
higher temperatures. The only temperature dependent factor in
$f_{tw}$ is the anisotropy and, indeed, a rapid decrease of
anisotropy near T$_c$ was found \cite{Zeh02,Bud02}. However, the
assumption that all other influences on the transition width are
temperature and field independent, are not appropriate to draw
exact conclusions on the temperature dependence of  the
anisotropy. Nevertheless, the increasing difference between the
onset ($B_{c2}$) and the offset (irreversibility field) of the
resistive transition with increasing magnetic field can be
explained by anisotropy and only minor corrections are needed to
obtain the field dependence of the transition width observed in
experiments.

In order to derive a model for the critical currents in
polycrystalline samples, grain boundaries are assumed to be the
dominant pinning centers. The dependence of the critical current
density on the applied magnetic field ($J_{c}\bot B$) is,
therefore, expected to be \cite{Dew74}
\begin{equation}
J_{c}\propto\frac{B_{c2}^{3/2}(\theta)(1-B/B_{c2}(\theta))^{2}}{\kappa^{2}(\theta)\sqrt{B}}
\propto \frac{(1-B/B_{c2}(\theta))^{2}}{\sqrt{B_{c2}(\theta)B}}
\label{equ3}
\end{equation}
for $B\leq B_{c2}(\theta)$ and zero otherwise. This equation was
originally developed for isotropic superconductors, but anisotropy
can be introduced through the angular dependence of the reversible
parameters. $\kappa=\kappa_{1}=B_{c2}/\sqrt{2}B_{c}$ was used. For
the sake of simplicity, we neglect any influence of the quality of
a specific grain boundary, the details of grain boundary pinning
as well as the angle between the local current and the applied
field, which may differ due to percolation. The main feature we
wish to address, is the disappearance of the critical current as
$(1-B/B_{c2}(\theta))^{2}$ at high fields and its rapid decrease
as $B^{-0.5}$ at low fields. With these critical current densities
of individual grains, we employ results on random resistor
networks to derive the critical current density of the whole
sample. There, the current in a network consisting of equal
resistors with a current voltage law $U \propto I^{\alpha}$ is
given near the percolation threshold by $I = \sigma_{eff}U^{1/
\alpha}$, with the conductivity $\sigma_{eff}\propto(p-p_{c})^{t}$
\cite{Str84} and $t=(d-1)\nu+(\zeta-\nu)/\alpha$ \cite{Ska74},
where $d$ denotes the dimensionality of the system, $\zeta$ is
about one and the correlation exponent $\nu$ is 0.88 in 3D systems
\cite{Gau83}. Most experimental data were obtained on systems
consisting of normal conducting ($\alpha=1$) and insulating
particles \cite{Abe75,Cle80,Lee86,Kir73}. In this case, the above
power law for $\sigma_{eff}$ was found to be an excellent
approximation and $t$ to be in reasonable agreement with the
theoretical prediction ($t=1.88$). The above dependence of $t$ on
$\alpha$ was confirmed numerically for 2D nonlinear resistor
networks \cite{Str84}. Its asymptotic ($\alpha\rightarrow\infty$)
value, $t=(d-1)\nu$, was proven for all dimensions \cite{Deu83}
($t=1.76$ for $d=3$). In the following we assume $\alpha$ to be
infinite and $t$ to be 1.76, which makes the current independent
of the voltage ("hard" superconductor), but the argument remains
essentially the same for any other $\alpha$ at fixed voltage. Note
that the influence of $\alpha$ on $t$ is very weak in 3D systems.
In a granular superconductor the conductivity $\sigma_{eff}$
becomes a function of the (local) current density $J$, if $p$ is
interpreted as the fraction of grains with critical current
densities larger than $J$. With the sample cross section
$\sigma_{0}$ and with an appropriate normalization to get correct
results for $p=1$, the conductivity (current) becomes
$I=\sigma_{eff}=\sigma_{0}(\frac{p(J)-p_{c}}{1-p_{c}})^{t}J=:\sigma_{p}J$.
We introduce $\sigma_{p}$, which can be interpreted as the total
cross section of all current paths existing at a certain fraction
$p$ of conducting particles. If the current is smaller than
$\sigma_{0}\min_{\theta}(J_{c}(B,\theta))$, $p$ is one and the
current flows homogeneously through the whole superconductor. Upon
enhancing the current by $\triangle I$, $p$ becomes smaller than
one and the current density in the remaining current paths
increases by $\triangle J=\triangle I/\sigma_{p}$. The remaining
capability of the grains to carry additional currents is reduced
by $\triangle J$. The current can be further enhanced until
$p=p_{c}$. Integrating over all infinitesimal small $\triangle I$
(or more easily $\triangle J$) and division through $\sigma_{0}$
leads to the macroscopic critical current density of the sample:
\begin{equation}
J_{c}(B)=\int\limits_{0}^{J_{c}^{max}(B)}(\frac{p(J)-p_{c}}{1-p_{c}})^{t}dJ
\label{equ4}
\end{equation}
where $J_{c}^{max}(B)$ is defined by $p(J_{c}^{max}(B))=p_{c}$. In
order to evaluate this expression numerically, 1001 grains were
chosen with $sin(\theta)$ equally distributed between 0 and 1.
Their critical currents were calculated according to (\ref{equ3})
and the probability distribution $p$ was calculated basically by
counting the number of grains with a critical current density
above a certain value. The integral was then replaced by a
summation over the discrete critical current densities of the
chosen grains. The model contains four parameters: the upper
critical field $B_{c2}$, the absolute value of the critical
current densities in (\ref{equ3}), the critical probability
$p_{c}$, and the anisotropy $\gamma$. While $B_{c2}$ can be
measured directly at high temperatures and extrapolated to low
temperatures, the remaining parameters are obtained by fitting the
logarithm of $J_{c}(B)$ obtained from (\ref{equ4}) to the
corresponding experimental data with the least squares method. Two
samples were used to compare experimental results with the
predictions of the model, a sintered bulk sample and a Cu-sheathed
powder-in-tube wire, prepared by an in-situ process \cite{Glo01}.
The samples were neutron irradiated in a fission reactor, in order
to change the reversible properties. Details of the irradiation
experiments can be found in \cite{Eis02a,Eis02b}. The upper
critical field of a similar bulk sample (from the same pellet) was
determined by the onset of the resistive transition
(Fig.~\ref{fig1}a) and the linear behavior below 30 K was
extrapolated to lower temperatures. For the wire, the onset of the
resistive transition cannot be evaluated for $B_{c2}$ due to the
low resistance of the copper sheath. Therefore, $B_{c2}$ was
obtained from SQUID measurements. The critical current densities
of the bulk sample were obtained from ac susceptibility
measurements, at high critical current densities by a technique
described in \cite{Eis00}, and near the irreversibility line by
evaluating the magnitude of the out-of-phase component in terms of
the Bean model. Since the first method is only applicable, if the
ac penetration depth is much smaller than the sample radius, and
the latter, if it is larger, intermediate current densities were
not evaluated. In the case of the wire, direct transport
measurements were made in liquid helium.

\begin{figure} \centering \includegraphics[width
= 0.8\columnwidth]{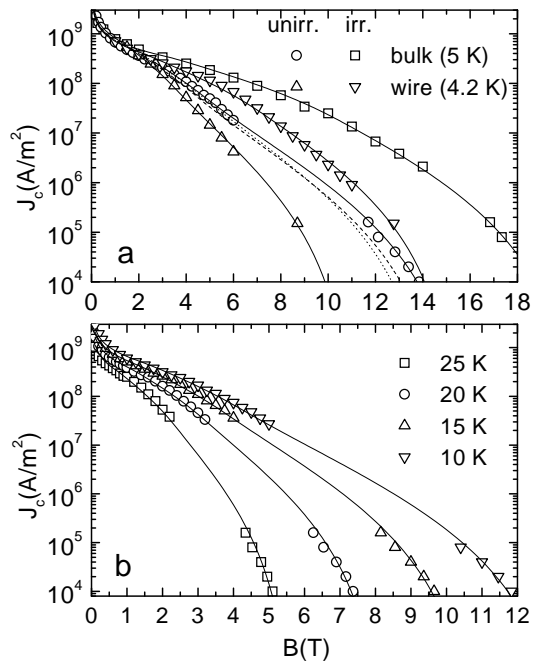} \caption{Critical current densities
of various samples at low temperatures (a) and of bulk MgB$_2$ at
higher temperatures (b). The solid lines represent the fit.}
\label{fig2}
\end{figure}

The results are plotted in Fig.~\ref{fig2}a together with the
fitted curves (solid lines). Since the excellent agreement between
experimental and calculated data (obtained by a three parameter
fit) might not be unique to the proposed model, it is most
important that the fit parameters (Table~\ref{tab1}) are indeed in
agreement with expectations. The anisotropy is found to be 4.4 for
both unirradiated samples. Literature data range from slightly
above 1 to 9 \cite{Buz01} and are obviously sample dependent. A
direct determination of the $B_{c2}$ anisotropy was made on single
crystals ($\gamma=4.6$ \cite{Zeh02}), aligned powders
($\gamma=1.7$ \cite{deL01}) and thin films ($\gamma=1.2-2$
\cite{Pat01}). For sintered polycrystalline samples (wires, bulk),
the anisotropy was extracted by a special technique from
magnetization measurements and estimated to be 6--7 \cite{Bud01}.
The large differences indicate that disorder affects anisotropy.
This is directly confirmed here by radiation-induced disorder
(Table~\ref{tab1}). The percolation threshold $p_{c}$ is similar
in the unirradiated samples (0.21 and 0.26). Theoretical
predictions for site percolation are about 0.2 for the face
centered cubic lattice (coordination number Z=12) and 0.31 for the
simple cubic lattice (Z=6) \cite{Mar97}. Although the grains in
our samples are not arranged in a regular lattice, our result
indicates that on average each grain is connected to 6--12
neighboring grains. This has been confirmed by SEM investigations.
Other experimental data range from 0.17 in pressed mixtures of
conducting spheres and Teflon powder \cite{Lee86}, to 0.31 in
systems consisting of silver coated and uncoated glass beads
\cite{Abe75} or 0.47 in granular W-Al$_{2}$O$_{3}$ films
\cite{Cle80}.

\begin{figure} \centering \includegraphics[width
= 0.7\columnwidth]{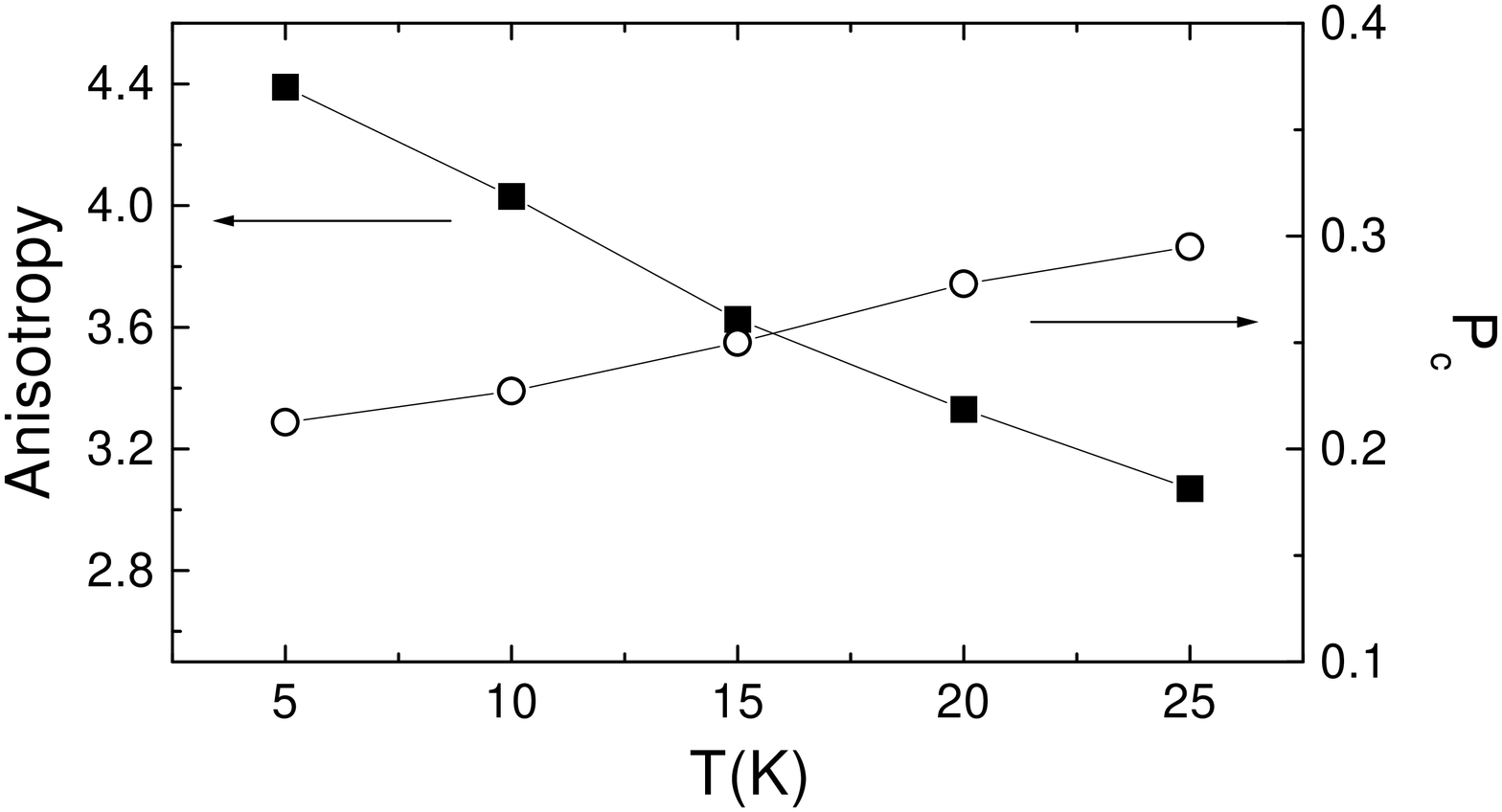} \caption{Temperature dependence of
the anisotropy and of the percolation threshold in bulk MgB$_2$.}
\label{fig3}
\end{figure}

Experimental data on the unirradiated bulk sample at higher
temperatures and the fitted $J_{c}(B)$ curves are plotted in
Fig.~\ref{fig2}b, the resulting values of $\gamma$ and $p_{c}$ in
Fig.~\ref{fig3}. While a temperature dependence of the $B_{c2}$
anisotropy was also found in single crystals \cite{Ang02,Zeh02}
and powders of MgB$_2$ \cite{Bud02}, the increase of $p_{c}$ with
increasing temperature is quite unexpected, since $p_{c}$ should
only depend on the arrangement of the grains within the sample. In
order to explain this temperature dependence, we investigate the
influence of $\gamma$ and $p_{c}$ on the critical currents. The
dotted line in Fig.~\ref{fig2}a represents the calculated current
densities, if $p_{c}$ of the unirradiated  bulk sample is enhanced
by 20 \%, leaving the other parameters constant. This has very
little effect for fields below $B_{c2}^{\bot}$. An increase of the
anisotropy by only 10 \% (dashed line in Fig.~\ref{fig2}a)
significantly decreases the critical currents to much lower
fields. The very small influence of $p_{c}$ compared to $\gamma$
at low fields makes the evaluation of the anisotropy quite robust
against changes in $p_{c}$. On the other hand, at a given
anisotropy, $p_{c}$ determines the irreversibility field and,
therefore, the transition width.  The fit parameter $p_{c}$ cannot
be interpreted only as the percolation threshold anymore, because
it also contains any other broadening effects on the transition.
Especially thermal fluctuations lead to an artificial increase of
the percolation threshold at high temperatures, but also material
inhomogeneities enhance $p_{c}$. The latter explains the increase
of $p_{c}$ after irradiation (Table~\ref{tab1}) due to the large
inhomogeneity of the defect structure following neutron
irradiation \cite{Eis02a}, which leads to a significant broadening
of the transition even at zero field ($\triangle T_{c}$). The wire
was irradiated in a cadmium cover \cite{Eis02b}, which reduces the
inhomogeneity of the resulting defect structure and the
enhancement of $p_{c}$.

Finally, the normalized transition width $f_{tw}$ was calculated
from the fit parameters of the unirradiated bulk sample
(Fig.~\ref{fig1}b: open circles) according to (\ref{equ2}). The
agreement with the data directly calculated from the measured
transition width is very satisfactory. We conclude that the width
of the resistive transition can be quantitatively explained under
realistic assumptions for the anisotropy (3.1--4.4) and for the
percolation probability (0.21--0.29), and that this contribution
is dominant at low temperatures.

In summary, we have shown that the dependence of the critical
currents in polycrystalline MgB$_2$ on the applied magnetic field
and the broadening of the resistive transition with increasing
magnetic fields can be explained by anisotropy and current
percolation. The rather low upper critical field $B_{c2}^{\bot}$,
turns out to be responsible for the strong field dependence of the
critical current densities in polycrystalline MgB$_2$.

\begin{table}
\caption{\label{tab1} Experimental data on the upper critical
fields ($B_{c2}^{\|}$) as well as anisotropy and percolation
threshold obtained from the fits of the critical current
densities.}
\begin{ruledtabular}
\begin{tabular}{lcccc}
&bulk (5 K)&bulk irr.&wire (4.2 K)&wire irr.\\ \hline
$B_{c2}^{\|}$ (extrapol.) &21.4 T&30 T&15.4 T&18.6 T\\
anisotropy&4.4&2.8&4.4&2.5\\ $B_{c2}^{\bot}$
($:=B_{c2}^{\|}/\gamma$)&4.9 T &10.7 T&3.5 T&7.4 T\\
$p_c$&0.21&0.4&0.26&0.29\\
\end{tabular}
\end{ruledtabular}
\end{table}

\end{document}